\documentclass[12pt,a4,epsf]{article} \usepackage{cite}
\usepackage{graphicx} 
\usepackage{amssymb} 
\global\arraycolsep=2pt
\input{epsf} 
\usepackage{graphicx}
\def\@fmsl@sh#1#2#3{\m@th\ooalign{$\hfil#1\mkern#2/\hfil$\crcr$#1#3$}}
 \def\eq#1\en{\begin{equation}#1\end{equation}}
\def\s[#1,#2]{[#1\stackrel{\star}{,}#2]}
\def\sx[#1,#2]{[#1\stackrel{\star_{x}}{,}#2]} 
\begin{document}
\makeatletter
\def\fmslash{\@ifnextchar[{\fmsl@sh}{\fmsl@sh[0mu]}}
\def\fmsl@sh[#1]#2{%
  \mathchoice
    {\@fmsl@sh\displaystyle{#1}{#2}}%
    {\@fmsl@sh\textstyle{#1}{#2}}%
    {\@fmsl@sh\scriptstyle{#1}{#2}}%
    {\@fmsl@sh\scriptscriptstyle{#1}{#2}}}
\def\@fmsl@sh#1#2#3{\m@th\ooalign{$\hfil#1\mkern#2/\hfil$\crcr$#1#3$}}
\makeatother

\thispagestyle{empty}
\begin{titlepage}
\begin{flushright}
CALT-68-2485\\
March 22, 2004
\end{flushright}

\vspace{0.3cm}
\boldmath
\begin{center}
  \Large {\bf Metric on a Statistical Space-Time}
\end{center}
\unboldmath
\vspace{0.8cm}
\begin{center}
  {\large Jacques Calmet\footnote{
email:calmet@ira.uka.de}$^*$ and Xavier Calmet\footnote{
email:calmet@theory.caltech.edu}$^\dagger$}\\ \end{center}
\begin{center}
{\sl $^*$Institute for Algorithms and Cognitive Systems (IAKS)\\
University of Karlsruhe (TH), D-76131 Karlsruhe, Germany} \\ {\sl
$^\dagger$California Institute of Technology, Pasadena, California
91125, USA}
\end{center}
\vspace{\fill}
\begin{abstract}
\noindent
We introduce a concept of distance for a space-time where the notion
of point is replaced by the notion of physical states e.g. probability
distributions. We apply ideas of information theory and compute the
Fisher information matrix on such a space-time. This matrix is the
metric on that manifold. We apply these ideas to a simple model and
show that the Lorentzian metric can be obtained if we assumed that the
probability distributions describing space-time fluctuations have
complex values. Such complex probability distributions appear in
non-Hermitian quantum mechanics.
\end{abstract}  
\end{titlepage}
\section{Introduction}

The concept of distance between two points plays a central role in any
physical theory. This notion is well defined as long as space-time can
be regarded as a classical system. Fundamental theories of nature are
quantum theories, i.e. the fields describing the particles and forces
have to be quantized or in other words replaced by operators. This
procedure is often referred to as second quantization. Whereas this
procedure is well understood for Yang-Mills theories that are relevant
to describe the electroweak and strong interactions, it is far less
clear how to quantize and how to renormalize Einstein's theory of
general relativity that describes gravitation. One thus expects that
general relativity has to be modified in the high energy regime. This
might be a hint that space-time structure is more complicated at very
short distances.

One common speculation is that at energies where fluctuations of the
metric become relevant space-time becomes fuzzy. There is then an
uncertainty in the measurement of a length, see e.g. \cite{Ng:2004qq}
for a recent review. It was noticed by Salecker and Wigner
\cite{Salecker} a long time ago that quantum mechanics and general
relativity considered together imply the existence of an uncertainty
in the measurement of a length
\begin{eqnarray}
\delta l^2 \ge \frac{\hbar l}{m c}
\end{eqnarray}
when a clock is used in a Gedanken experiment to measure a distance.

We will make a different and more radical assumption. It has been
proposed long ago by Rosen \cite{rosen} that the notion of point might
not be relevant anymore at short distances or equivalently at high
energies. The basic assumption is that a physical point is not a well
localized entity but is fuzzy in the sense that the only information
one has is of statistical nature, namely that the ``mathematical''
point is localized within a certain volume. We substitute the notion
of a four dimensional point $x^\mu$ by the notion of a distribution
$\bar x^\mu=p_{\theta^\mu}(x^\mu)$. Such an assumption is not that
different from the one made in noncommutative geometry where points
$x^\mu$ are replaced by noncommutative operators $\hat x^\mu$
\cite{Snyder:1946qz}. One can imagine different concepts for
distance. For example, in the case of noncommutative geometry, Connes'
distance \cite{Connes:2000ti} can be used. The question we want to
address is the following: how can a distance be defined if the notion
of a well localized point is replaced by the notion of a distribution?
It turns out that ideas from information theory can be applied. We
will propose a definition for the distance on such a statistical
space-time. This paper is organized as follows: we will first review
the concept of Shannon entropy and explain how it leads to the
introduction of Fisher's metric. We will then apply these ideas to a
simple model for a fluctuating space-time, define a distance and
compute the metric on the manifold of distributions. We then conclude.

\section{Brief review of Fisher information metric}

There are many excellent reviews and books on Fisher information
metric, a nice introduction can be found in \cite{Frieden}. A distance
$d(P_1,P_2)$ between two points $P_1$ and $P_2$ has to satisfy the
following three axioms:
\begin{itemize}
\item[1.] Positive definiteness: $\forall P_1, P_2: d(P_1,P_2) \ge 0$
\item[2.] Symmetry $d(P_1,P_2)=d(P_2,P_1)$
\item[3.] Triangle inequality: $\forall P_1, P_2, P_3: d(P_1,P_2) \le
d(P_1,P_2) +d(P_1,P_3)$.
\end{itemize}
This concept of distance can be traced back to Aristotle  and Euclid.

It is often useful to introduce a concept of distance between elements
of a more abstract set.  For example, one could ask what is the
distance between two distributions between e.g. the Gaussian and
binomial distributions. It is useful to introduce the concept of
entropy as a mean to define distances. In information theory,
Shannon entropy \cite{Shannon} represents the information content of
a message or, from the receiver point of view, the uncertainty about
the message the sender produced prior to its reception. It is defined
as
\begin{eqnarray}
- \sum_i p(i) \log p(i),
\end{eqnarray}
where $p(i)$ is the probability of receiving the message $i$. The unit
used is the bit. The relative entropy can be used to define a
``distance'' between two distributions $p(i)$ and $g(i)$. The
Kullback-Leibler \cite{Kullback} distance or relative entropy is defined as
\begin{eqnarray}
D(g||p)&=& \sum_i g(i) \log \frac{g(i)}{p(i)}
\end{eqnarray}
where $p(i)$ is the real distribution and $g(i)$ is an assumed
distribution. Clearly the Kullback-Leibler relative entropy is not a
distance in the usual sense: it satisfies the positive definiteness
axiom, but not the symmetry or the triangle inequality axioms. It is
nevertheless useful to think of the relative entropy as a distance
between distributions.

The Kullback-Leibler distance is relevant to discrete sets. It can be
generalized to the case of continuous sets. For our purposes, a
probability distribution over some field (or set) $X$ is a
distribution $p:X \in \mathbb{R}$, such that
\begin{itemize}
\item[1.] $\int_X d^4\!x \  p(x)=1$
\item[2.] For any finite subset $S\subset X$, $\int_S d^4\!x \ p(x)>0$.
\end{itemize}
We shall consider families of distributions, and parameterize them by a
set of continuous parameters $\theta^i$ that take values in some open
interval $M \subseteq \mathbb{R}^4$. We use the notation $p_\theta$ to
denote members of the family. For any fixed $\theta$, $p_\theta: x
\mapsto p_\theta(x)$ is a mapping from $X$ to $\mathbb{R}$. We shall
consider the extension of the family of distributions $F=\{ p_\theta|
\theta \in M\}$, to a manifold ${\cal M}$ such that the points $p\in
{\cal M}$ are in one to one correspondence with the distributions
$p\in F$. The parameters $\theta$ of $F$ can thus be used as
coordinates on ${\cal M}$.

The Kullback number is the generalization of the Kullback-Leibler
distance for continuous sets. It is defined as
\begin{eqnarray}
I(g_\theta|| p_\theta)&=&  
\int d^4\!x  g_\theta(x) \log \frac{g_\theta(x)}{p_\theta(x)}.
\end{eqnarray}
Let us now study the case of an infinitesimal difference between
$q_\theta(x)=p_{\theta +\epsilon v}(x)$ and $p_\theta(x)$:
\begin{eqnarray}
I(p_{\theta +\epsilon v}||p_\theta)&=&  \int d^4\!x  p_{\theta +\epsilon v}(x) 
\log \frac{p_{\theta +\epsilon v}(x)}{p_\theta(x)}.
\end{eqnarray}
Expanding in $\epsilon$ and keeping $\theta$ and $v$ fix one finds
(see e.g. \cite{rodriguez2,rodriguez3}):
\begin{eqnarray}
I(p_{\theta +\epsilon v}||p_\theta)&=&I(p+\epsilon||p)
\vert_{\epsilon=0}+ \epsilon \ I^\prime(\epsilon)\vert_{\epsilon=0}+
\frac{1}{2}\epsilon^2 \
I^{\prime\prime}(\epsilon)\vert_{\epsilon=0}+{\cal O}(\epsilon^3).
\end{eqnarray}
One finds $I(0)=I^\prime(0)=0$ and
\begin{eqnarray}
I^{\prime\prime}(0)&=& v^\mu \left ( \int_X d^4\!x p_\theta(x) \left 
(\frac{1}{p_\theta(x)} \frac{\partial p_\theta(x)}{\partial \theta^\mu} \right )
 \left  
(\frac{1}{p_\theta(x)} 
\frac{\partial p_\theta(x)}{\partial \theta^\nu} \right ) \right ) v^\nu.
\end{eqnarray}
We can now identify the Fisher information metric \cite{Fisher} on a
manifold of probability distributions as
\begin{eqnarray}
g_{\mu\nu}=\int_X d^4\!x p_\theta(x) \left 
(\frac{1}{p_\theta(x)} \frac{\partial p_\theta(x)}{\partial \theta^\mu} \right )
 \left  
(\frac{1}{p_\theta(x)} 
\frac{\partial p_\theta(x)}{\partial \theta^\nu} \right ).
\end{eqnarray}
It has been show that this matrix is a metric on a manifold of
probability distributions, see e.g.  \cite{rodriguez1}. Corcuera and
Giummol\`e \cite{Corcuera} have shown that the Fisher information
metric is invariant under reparametrization of the sample space $X$
and that it is covariant under reparametrizations of the manifold,
i.e. the parameter space, see e.g. \cite{Wagenaar} for a
review. Fisher's information matrix plays an important role in many
different fields. This concept appears in such different fields as
e.g.  instanton calculus \cite{Yahikozawa:2003ij}, ontology
\cite{jcalmet} or econometrics \cite{Marriott}. Symbolic computations
of Fisher information matrices have also been considered
\cite{Peeters}.

\section{Fisher information metric and distance on fluctuating spaces}

Let us now apply the ideas developed in the previous chapter to a
simple model of space-time. Let us assume that the notion of points
$x^\mu$ is replaced by a state $\bar x$ that could be for example a
distribution $p_{\theta^\mu}(x^\mu)$. We propose the following
definition
\begin{eqnarray}
I(q_{{\theta'}^\mu}(x^\mu)||p_{\theta^\mu}(x^\mu) )&=&  
\int d^4\!x   q_{{\theta'}^\mu}(x^\mu)\log \frac{q_{{\theta'}^\mu}(x^\mu)}{p_{\theta^\mu}(x^\mu)}
\end{eqnarray}
for the distance between two ``points'' $p_{\theta^\mu}(x^\mu)$ and
$q_{{\theta'}^\mu}(x^\mu)$. The metric on the manifold of
distributions is given locally by
\begin{eqnarray}
g_{\mu\nu}=\int_X d^4\!x p_\theta(x) \left 
(\frac{1}{p_\theta(x)} \frac{\partial p_\theta(x)}{\partial 
\theta^\mu} \right )
 \left  
(\frac{1}{p_\theta(x)} 
\frac{\partial p_\theta(x)}{\partial \theta^\nu} \right )
\end{eqnarray}
and corresponds to the Fisher information matrix.  The distance
between two points $A^\mu$ and $B^\nu$ on the manifold is given by
$d(A^\mu, B^\nu)=\sqrt{g_{\mu\nu}A^\mu B^\nu}$.

As a example, one can consider for example a 3-dimensional Gaussian
distribution
\begin{eqnarray}
p_\theta(x)=\frac{1}{(2 \pi a^2)^{\frac{2}{3}}} 
\exp\left(-\frac{(x-\theta_1)^2+(y-\theta_2)^2+(z-\theta_3)^2}{2 a^2}\right)
\end{eqnarray}
the Fisher metric reads $g_{ij}=1/a^2 \mbox{diag}(1,1,1)$, note that
one has the freedom to rescale the relative entropy by a factor $a^2$,
it wish case it is simply the matrix $\mbox{diag}(1,1,1)$. The Fisher
information matrix as already been calculated in the literature for a
Gaussian distribution, see e.g.  \cite{Caticha}, where the parameter
$a$ was interpreted as $\theta_0$, in our case we choose to treat $a$
as scale parameter of the model and not to identify it with a
coordinate on the manifold.

An interesting question arises: can we generate a four dimensional
space-time with a Lorentzian signature diag$(-1,1,1,1)$? One has to
solve the following system of equations:
\begin{eqnarray}
g_{00}&=&\int d^4\!x \frac{1}{p_\theta(x)}\left 
( \frac{\partial p_\theta(x)}{\partial 
\theta^0} \right )^2 =-1 \\ \nonumber
g_{ii}&=&\int d^4\!x \frac{1}{p_\theta(x)}\left 
( \frac{\partial p_\theta(x)}{\partial 
\theta^i} \right )^2 =1 \ \mbox{for} \ i \in \{1,2,3\} \\ \nonumber
g_{\mu\nu}&=&\int d^4\!x \frac{1}{p_\theta(x)}\left 
( \frac{\partial p_\theta(x)}{\partial 
\theta^\mu} \right ) \left ( \frac{\partial p_\theta(x)}{\partial 
\theta^\nu} \right )=0 \ \mbox{for} \ \mu \ne  \nu.
\end{eqnarray}
The first of these equations $g_{00}=-1$ is not solvable if
$p_\theta(x)$ is a real probability distribution as the ones usual
consider in quantum mechanics. In that case $\frac{1}{p_\theta(x)}$ is
always positive and $\left( \frac{\partial p_\theta(x)}{\partial
\theta^0} \right )^2$ is also positive. One way out is to extend the
definition we gave of a probability distribution to include complex
probability distributions. This is not as surprising as it might
sound. Non-Hermitian quantum mechanics has been used to deal with
physical phenomena involving metastable finite-lifetime states,
so-called resonances \cite{Moiseyev} and for the study of
delocalization phenomena such as bacteria populations, vortex
spinning in superconductors or hydrodynamical problems
\cite{Hatano}. In non-Hermitian quantum mechanics density
probabilities are complex functions \cite{Barkay}. The complex
transition probability is a measurable quantity in
e.g. electron-quantum dot scattering-like experiments \cite{schuster}.

As an ansatz, let us consider the complex probability distribution:
\begin{eqnarray}
p_\theta(x)=\frac{1}{(2 \pi a^2)^{2}} 
\exp \left (-\frac{(t-i\theta_0)^2+(x-\theta_1)^2+(y-\theta_2)^2+(z-\theta_3)^2}{2 a^2}\right).
\end{eqnarray}
This distribution is normalized: $\int d^4\!x p_\theta(x)=1$ and leads
to the following Fisher information matrix
\begin{eqnarray}
(g_{\mu\nu})=\left (\begin{array}{cccc}
-1 & 0 & 0 & 0\\
0 & 1 & 0 & 0\\
0 & 0 &1  & 0\\
0 & 0 & 0 & 1\\
\end{array}\right),
\end{eqnarray}
after a rescaling by $a^2$.  It is interesting to note that a complex
Gaussian distribution can lead to a Lorentzian metric. This might be a
hint that the Lorentzian metric is due to a non-Hermitian nature of
quantum gravity, whatever that theory might be. In our case the
signature of space-time and in particular the special nature of time
can be related to an imaginary parameter $i\theta_0$.  The Fisher
information metric can be seen as a statistical average over
space-time fluctuation. The macroscopic Lorentzian metric appears as a
consequence of the statistical distribution of space-time points.

\section{Conclusions}
In this paper we have considered a new concept of distance for a
space-time where points are replaced by states that can be for example
distributions or operators. We apply ideas developed in information
theory, the Fisher information is the metric on the manifold of
states.  We define the distance between two vectors on this manifold.
We apply this new concept to a simple model for a fluctuating
space-time. We show that if we extend the domain of definition of the
probability distributions from real to complex numbers, we can recover
a Lorentzian metric.


\end{document}